\documentclass[prd,floatfix,preprintnumbers,letterpaper]{revtex4}
\usepackage{amsmath,amssymb,graphicx}
\usepackage{subfigure}
\usepackage{hyperref}

\begin{document}

\title{Several parametrization dark energy models comparison with Statefinder hierarchy}

\author{Jing-Zhao Qi}
\author{Wen-Biao Liu}
\email{wbliu@bnu.edu.cn}
\affiliation{Department of Physics, Institute of Theoretical Physics, Beijing Normal University, Beijing, 100875, China}

\begin{abstract}
We employ the Statefinder hierarchy and the growth rate of matter perturbations to explore the discrimination of $\Lambda$CDM and some parametrization dark energy models including the Chevallier-Polarski-Linder (CPL), the Jassal-Bagla-Padmanabhan (JBP), the Pad\'{e}(\uppercase\expandafter{\romannumeral1}), (\uppercase\expandafter{\romannumeral2}). We find that the statefinder $S_3^{(m)}$ containing third derivatives of $a(t)$ can differentiate CPL and JBP from $\Lambda$CDM and Pad\'{e}(\uppercase\expandafter{\romannumeral1}), (\uppercase\expandafter{\romannumeral2}). While the statefinder $S_4^{(1)}$ involving fourth order derivatives of $a(t)$ has more powerful discrimination that it can distinguish the Pad\'{e}(\uppercase\expandafter{\romannumeral1}), (\uppercase\expandafter{\romannumeral2}) from $\Lambda$CDM. In addition, we show that the growth rate of matter perturbations does not play a significant role for discrimination of such parametrization dark energy models.
\end{abstract}

\maketitle

\section{Introduction}  \label{introduction}
The late time cosmic acceleration has been supported by many independent cosmological observations, including the type Ia supernovae (SNIa) \citep{riess1998supernova}, large scale structure \citep{tegmark2004cosmological}, cosmic microwave background (CMB) anisotropy \citep{spergel2003wmap} etc. Dark energy is believed to have unveiled the mystery of the cosmic acceleration, via the ratio of its pressure and energy density as well as the equation of state (EoS) parameter $w<-1/3$. Depending whether EoS is constant or time-dependent, the dark energy candidates can be divided into many subclasses. The cosmological constant model ($\Lambda$CDM) with $w=-1$ is the most robust model but with the fine-tuning problem \citep{weinberg1989cosmological,weinberg2000cosmological} and coincidence problem \citep{1999PhRvL..82..896Z}. This led to a widespread speculation that the EoS of vacuum energy is not a constant. And then, a number of dynamic dark energy models have been proposed, such as quintessence \citep{Caldwell:2005tm,Zlatev:1998tr,Tsujikawa:2013fta}, K-essence \citep{chiba2000kinetically,armendariz2000dynamical}, phantom \citep{kahya2007quantum,onemli2004quantum,singh2003cosmological}, Chaplygin gas \citep{bento2002generalized,kamenshchik2001alternative}, and so on. On the other hand, the parameterization for the EoS of dark energy is a useful tool, which has been widely employed to analyse the behavior of dark energy \citep{riess2004type,barboza2009generalized,zhang2015exploring,maor2001limitations,
chevallier2001accelerating,linder2003exploring,Jassal:2004ej,Wei:2013jya}.

Facing more and more dark energy models, it is significant to discriminate various models. \citet{Sahni:2002fz} proposed a geometrical diagnostic called Statefinder to distinguish numerous dark energy models, such as $\Lambda$CDM, quintessence \citep{Sahni:2002fz,shojai2009statefinder,zhang2005statefinder}, phantom \citep{bao2007statefinder,chang2007statefinder}, holographic dark energy model \citep{zhang2005statefinder}, Ricci dark energy model \citep{feng2008statefinder}, and so on. Statefinder is a pair parameters $\{r,s\}$ containing the third derivative of the scale factor, $a^{(3)}/aH^3$. Different models will show different evolutionary trajectories in $r-s$ plane. In addition, another diagnostic named $Om(z)$ \citep{sahni2008two}, constructed from the Hubble parameter, can also be employed to distinguish the dark energy models. Recently, \citet{Arabsalmani:2011fz} introduced a more refined diagnostic called as 'Statefinder hierarchy', which involves high order derivative of the scale factor, $a^{(n)}/aH^n$, $n\geq3$. Statefinder hierarchy has a greater ability to distinguish various cosmological models. Sometimes, some dark energy models could not be discriminated by original Statefinder \citep{cui2014comparing}, but the hierarchy could break their degeneracy \citep{zhang2014diagnosing}. In addition, the growth rate of matter perturbations \citep{Arabsalmani:2011fz} also could be a supplement for Statefinder to discriminate cosmological models.

In this paper, we focus on the parametrizations for the EoS of dark energy. So far, there are a large number of parametrizations in the literature. Whether they could be distinguished will be investigated. We use Statefinder hierarchy and the growth rate of matter perturbations to compare some previously well-studied parametrizations: the Chevallier-Polarski-Linder (CPL), the Jassal-Bagla-Padmanabhan (JBP) and Pad\'{e}(\uppercase\expandafter{\romannumeral1}), (\uppercase\expandafter{\romannumeral2}) parametrizations whose proposal are based on the Pad\'{e} approximant \citep{Wei:2013jya}.

In Sec. \ref{eos}, CPL, JBP, Pad\'{e}(\uppercase\expandafter{\romannumeral1}) and Pad\'{e}(\uppercase\expandafter{\romannumeral2}) parametrizations for the EoS  will be introduced. The Statefinder hierarchy and the growth rate of matter perturbations are briefly reviewed in Sec. \ref{state}. In Sec. \ref{compare}, distinguishing parametrization dark energy models with the Statefinder hierarchy and the growth rate of matter perturbations will be presented. The last section will give some conclusions and discussions.

\section{Parametrizations for the EoS} \label{eos}
So far, there are a large number of parametrizations in the literature. Among them, the most popular one is the Chevallier-Polarski-Linder (CPL) parametrization \citep{chevallier2001accelerating,linder2003exploring}:
\begin{equation}
w_{de}=w_0+w_a(1-a)=w_0+w_a\frac{z}{1+z}, \label{cpl}
\end{equation}
where $w_0$ and $w_a$ are constants. It is obvious that at the present epoch $w|_{z=0}=w_0$ and in the early time $w(z\gg 1)\sim w_0+w_a$. Its normalized Hubble parameter for a flat universe is
\begin{eqnarray}
E^2(z)&=&\frac{H^2(z)}{H^{2}_0} =\Omega_{m0}(1+z)^3+(1-\Omega_{m0})(1+z)^{3(1+w_0+w_a)}\exp\left(-3w_az/(1+z)\right). \label{cplE}
\end{eqnarray}
According to Nine-year Wilkinson Microwave Anisotropy Probe (WMAP) observations \citep{hinshaw2013nine}, the values of parameters we take are: $\Omega_{m0}=0.288$, $w_0=-1.17$ and $w_a=0.35$.

Jassal-Bagla-Padmanabhan (JBP) parametrization \citep{Jassal:2004ej} is a variant of CPL parametrization
\begin{eqnarray}
w_{de}=w_0+w_a\frac{z}{(1+z)^2}. \label{JBP}
\end{eqnarray}
The parameter $w_0$ is dominant at both low and high redshifts: $w(z=0)=w_0$ and $w(z\gg 1)\sim w_0$. The normalized Hubble parameter is
\begin{eqnarray}
E^2(z)&=&\Omega_{m0}(1+z)^3+(1-\Omega_{m0})(1+z)^{3(1+w_0)}\exp\left(3w_az^2/2(1+z)^2\right). \label{jpbE}
\end{eqnarray}
The best-fit values of parameters we take are: $\Omega_m=0.288$, $w_0=-1.21$ and $w_a=1.28$ \citep{vazquez2012reconstruction}.

In fact, the above two models could be regarded as the Taylor series expansion or variant, while the Pad\'{e} approximant often gives better approximation \citep{PP}. \citet{Wei:2013jya} introduced two parametrizations for the EOS of dark energy based on the Pad\'{e} approximant, and confronted them with the observational data. In addition, some advantages of these EoS parametrizations have been demonstrated, such as they could work well at $z\sim -1$ where CPL and JBP diverge. The type (\uppercase\expandafter{\romannumeral1}) Pad\'{e} parametrization is
\begin{eqnarray}
w_{de}=\frac{w_0+(w_0+w_a)}{1+(1+w_b)z}, \label{pp1}
\end{eqnarray}
where $w_0$, $w_a$ and $w_b$ are constants. In the flat FRW universe, its normalized Hubble parameter is
\begin{eqnarray}
E^2(z)&=&\Omega_{m0}(1+z)^3+(1-\Omega_{m0})(1+z)^{3(1+w_0+w_a+w_b)/(1+w_b)}\left(1+\frac{w_bz}{1+z}\right)^{-3(w_a-w_0w_b)/[w_b(1+w_b)]}. \label{pp1E}
\end{eqnarray}
The values of parameters are constrained from observational data as: $\Omega_{m0}=0.280$, $w_0=-0.995$, $w_a=-0.02$, and $w_b=-0.052$ \citep{Wei:2013jya}.

The type (\uppercase\expandafter{\romannumeral2}) Pad\'{e} parametrization has been proposed as
\begin{eqnarray}
w_{de}=\frac{w_0+w_1\ln{a}}{1+w_2\ln{a}}, \label{pp2}
\end{eqnarray}
where $w_0$, $w_1$ and $w_2$ are all constants. Its normalized Hubble parameter is
\begin{eqnarray}
E^2(z)&=&\Omega_{m0}(1+z)^3+(1-\Omega_{m0})(1+z)^{3(w_1+w_2)/w_2}\left[1-w_2\ln{(1+z)}\right]^{3(w_1-w_0w_2)/w_2^2}. \label{pp2E}
\end{eqnarray}
The best-fit values of parameters are given as: $\Omega_{m0}=0.280$, $w_0=-0.996$, $w_1=0.200$, and $w_2=-0.139$ \citep{Wei:2013jya}.

\section{The statefinder hierarchy and the growth rate of perturbations}\label{state} \label{statefinder}
\subsection{The statefinder hierarchy}
Now, we adopt the assumption of a flat and homogeneous Friedmann-Robertson-Walker spacetime. In this case, the scale factor $a(t)$ describes the dynamics of universe, which can be Taylor expanded around the present time $t_0$ as follows
\begin{eqnarray}
\frac{a(t)}{a_0}=1+\sum _{n=1}^{\infty} \frac{A_n(t_0)}{n!}\left[H_0(t-t_0)\right]^n, \label{Taylor}
\end{eqnarray}
where
\begin{eqnarray}
A_n \equiv \frac{a^{(n)}}{aH^n}, ~~~~~ n \in N, \label{An}
\end{eqnarray}
$a^{(n)}$ is the $n$th derivative of the scalar factor $a(t)$ with respect to time. In fact, $-A_2$ is called deceleration factor $q$. $A_3$ is the jerk "$j$" \citep{Visser:2003vq} as well as the famous Statefinder "$r$" \citep{Sahni:2002fz} which is used to discriminate cosmological models \citep{Gao:2010ia,Panotopoulos:2007zn,Zhang:2007uh,Wei:2007zs}. $A_4$ is the snap "$s$" and $A_5$ is referred to as the lerk "$l$" \citep{Dabrowski:2005fg}. For the $\Lambda$CDM, all the $A_n$ parameters could be expressed as the functions of the deceleration parameter $q$, or the matter density parameter $\Omega_m$.
\begin{eqnarray}
A_2&=&1-\frac{3}{2}\Omega_m,\nonumber \\
A_3&=&1,\nonumber \\
A_4&=&1-\frac{3^2}{2}\Omega_m, ~~~ etc, \label{Anl}
\end{eqnarray}
where $\Omega_m=\frac{2}{3}(1+q)$ and $\Omega_m=\Omega_{m0}(1+z)^3/E^2(z)$ for $\Lambda$CDM. While the general forms of Eqs. (\ref{An}) could be rewritten as
\begin{eqnarray}
A_2&=&-q=1-(1+z)\frac{1}{E}\frac{dE}{dz},\nonumber \\
A_3&=&(1+z)\frac{1}{E^2}\frac{d[E^2(1+q)]}{dz}-3q-2,\nonumber  \\
A_4&=&\frac{-(1+z)}{E^3}\frac{d[E^3(2+3q+A_3)]}{dz}+4A_3+3q(q+4)+6.
\end{eqnarray}
And then, \citet{Arabsalmani:2011fz} proposed the \textit{Statefinder hierarchy} $S_n$ which can be defined as \citep{Li:2014mua}

\begin{eqnarray}
S_2&=&A_2+\frac{3}{2}\Omega_m,\nonumber \\
S_3&=&A_3,\nonumber \\
S_4&=&A_4+\frac{3^2}{2}\Omega_m, ~~~~~etc. \label{Sn}
\end{eqnarray}
Obviously, the Statefinder hierarchy is always equal to $1$ for $\Lambda$CDM during the entire evolution of the universe
\begin{equation}
S_n|_{\Lambda CDM}=1.
\end{equation}
These equations make it possible to define a null diagnostic for $\Lambda$CDM, since for evolving dark energy models some equalities of $S_n$ may be deviate from unity, which exactly could be a diagnostic tool.

For $n\geq 3$, using the relationship $\Omega_m=(2/3)(1+q)$ which is valid in $\Lambda$CDM, we can define an alternate form of the Statefinders
\begin{eqnarray}
S_3^{(1)}&:=&S_3, \nonumber \\
S_4^{(1)}&:=&A_4+3(1+q), ~~~~~etc. \label{S41}
\end{eqnarray}
$S_n^{(1)}$ is similar to $S_n$, as $S_n^{(1)}|_{\Lambda CDM}=1$. And then, it also can be a diagnostic. Certainly, the pair \{$S_n,S_n^{(1)}$\} is expected to differentiate varieties of cosmological models.

Except for the parameter $r$, the original Statefinder constructed another parameter $s\equiv \frac{r-1}{3(q-1/2)}$ \citep{Sahni:2002fz}. In analogy with $s$, \citet{Li:2014mua} defined the second member of the Statefinder hierarchy as follows
\begin{eqnarray}
S_n^{(2)}=\frac{S_n^{(1)}-1}{\alpha(q-1/2)}, \label{Sn2}
\end{eqnarray}
where $\alpha$ is a constant. That is to say, the original Statefinder pair $\{r,s\}$ is $\{S_3^{(1)},S_3^{(2)}\}$ with $\alpha=3$. Corresponding, the $S_4^{(2)}$ can be defined as
\begin{eqnarray}
S_4^{(2)}=\frac{S_4^{(1)}-1}{9(q-1/2)}. \label{S42}
\end{eqnarray}
For $\Lambda$CDM, $\{S_n^{(1)},S_n^{(2)}\}=\{1,0\}$, and then pairs $\{S_n,S_n^{(2)}\}$ and $\{S_n^{(1)},S_n^{(2)}\}$ are looking forward to be useful.

\subsection{The growth rate of matter perturbations}
The fractional growth parameter $\epsilon(z)$ \citep{Acquaviva:2010vr,Acquaviva:2008qp} could be a supplement for the Statefinders, which is defined as
\begin{eqnarray}
\epsilon(z):=\frac{f(z)}{f_{\Lambda CDM}(z)}, \label{ez}
\end{eqnarray}
where $f(z)=d\ln{\delta}/d\ln{a}$ is the growth rate of linearized density perturbations. It could be parameterized as \citep{Wang:1998gt}
\begin{eqnarray}
f(z)&\simeq& \Omega_m^\gamma(z), \label{f} \\
\gamma(z)&\simeq&\frac{3}{5-\frac{w}{1-w}}+\frac{3}{125}\frac{(1-w)(1-\frac{3}{2}w)}{(1-\frac{6}{5})^3}(1-\Omega_m(z)),
\end{eqnarray}
where $w$ is the EoS parameter of dark energy. If $w$ is a constant or varies slowly with time, the above approximation will work reasonably well. The Ref. \citep{Arabsalmani:2011fz} has demonstrated that the fractional growth parameter $\epsilon(z)$ can be used combining with the Statefinder hierarchy to define a composite null diagnostic: \{$\epsilon(z),S_n$\} or \{$\epsilon(z),S_n^{(m)}$\}. Indeedly, it is quite effective, as DPG, $\omega$CDM, and $\Lambda$CDM have been differentiated in Ref. \citep{Arabsalmani:2011fz}. For $\Lambda$CDM, $\gamma \simeq 0.55$ \citep{Wang:1998gt}, and $\epsilon(z)=1$, so \{$\epsilon(z),S_n$\}=\{1,1\}.

\section{Distinguishing parametrization dark energy models with statefinder hierarchy}\label{compare}
\begin{figure}
\centering
\includegraphics[width=8cm,height=6cm]{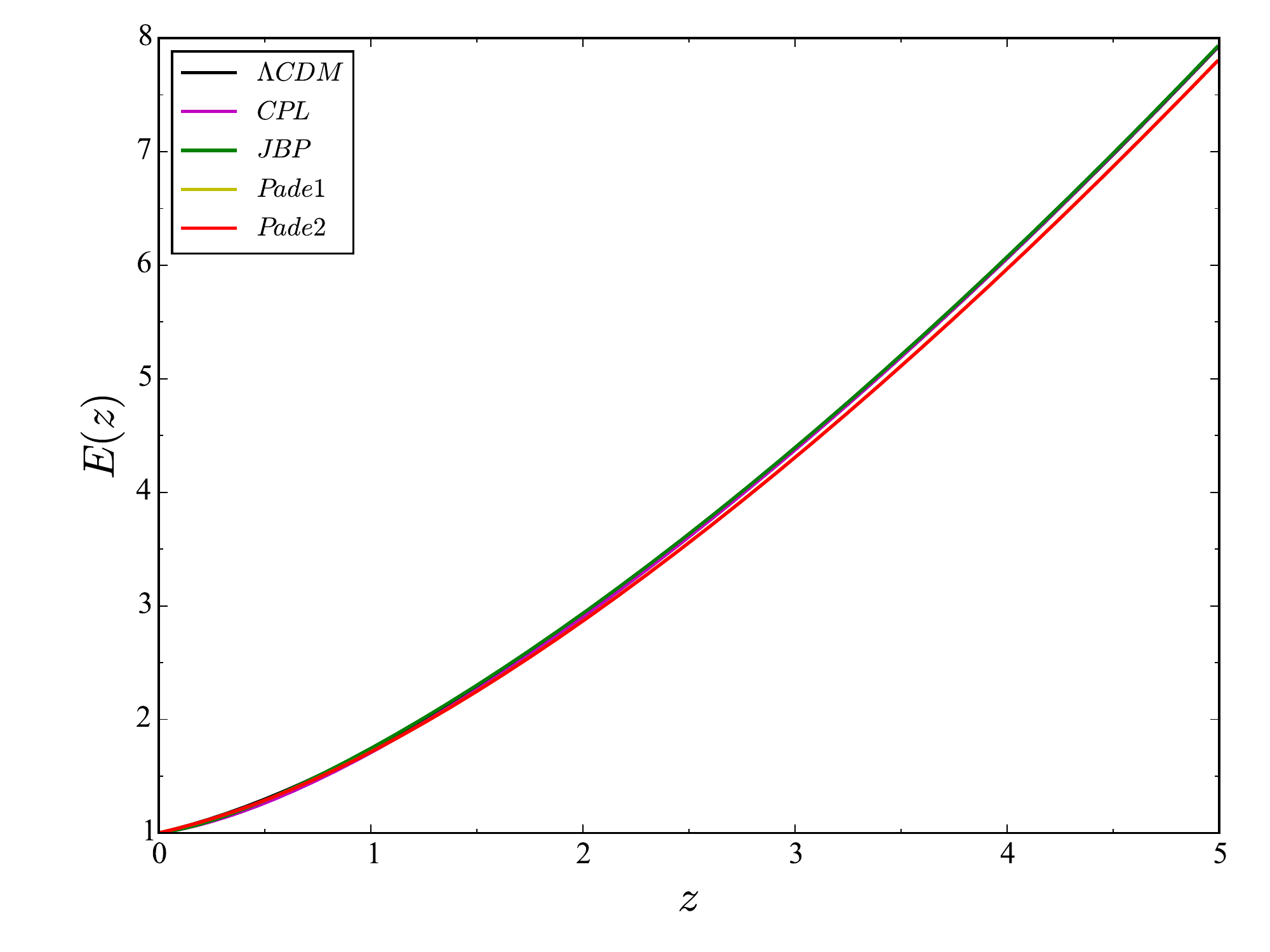}
\includegraphics[width=8cm,height=6cm]{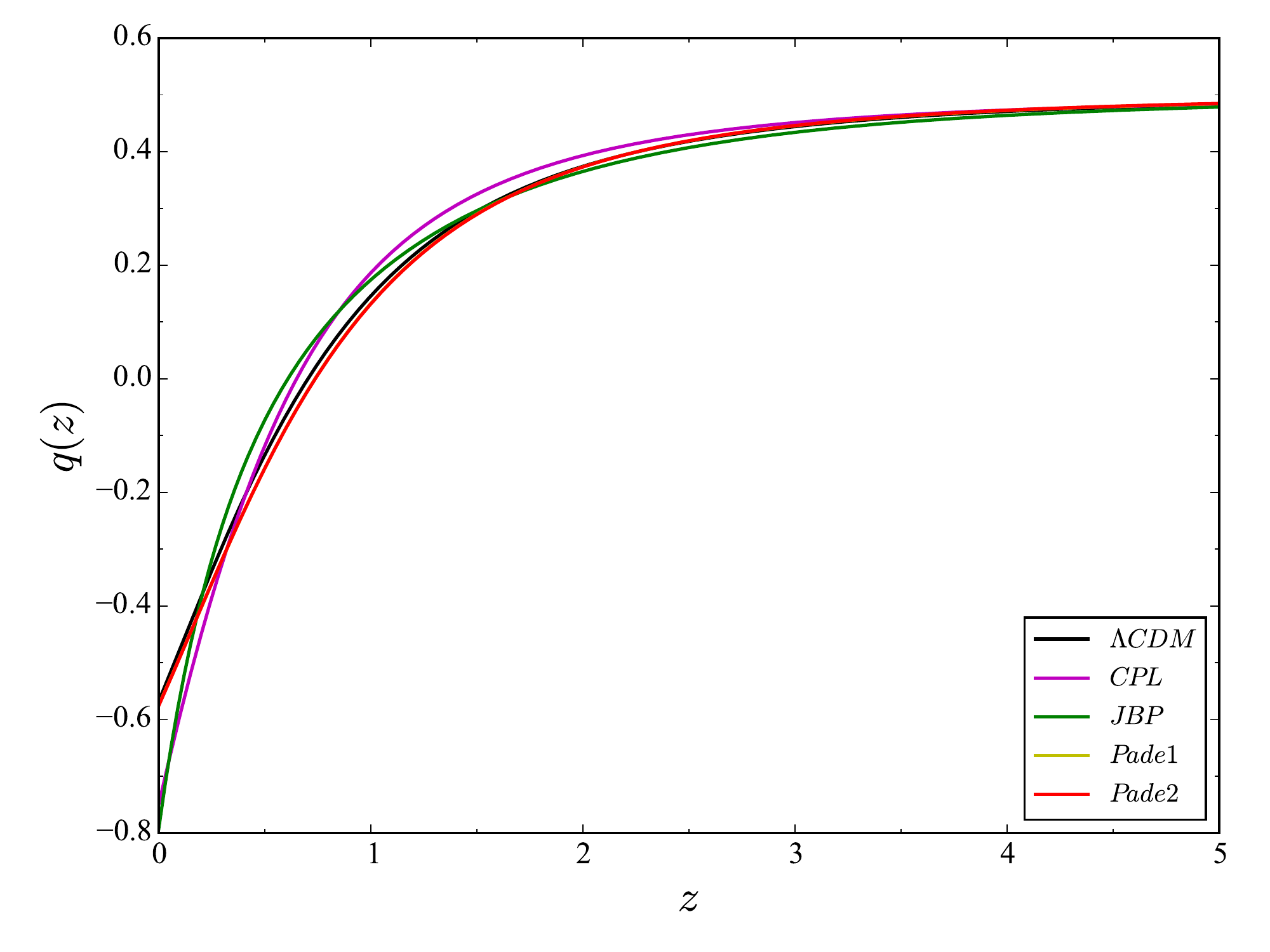}
\caption{The evolutions of the $E(z)$ and $q(z)$, respectively, versus redshift $z$.}\label{F1}
\end{figure}

Fig. \ref{F1} displays the evolutions of the $E(z)$ and $q(z)$ for $\Lambda$CDM, CPL, JBP, Pad\'{e}(\uppercase\expandafter{\romannumeral1}), Pad\'{e}(\uppercase\expandafter{\romannumeral2}). It is shown that these models can not be distinguished completely. In other words, they are highly degenerate. Therefore, the current observational data for $E(z)$ could not distinguish or eliminate these cosmological models. We expect that the Statefinder hierarchy takes advantage of its powerful discrimination to break the degeneracy of these models.

\begin{figure}
\centering
\includegraphics[width=8cm,height=6cm]{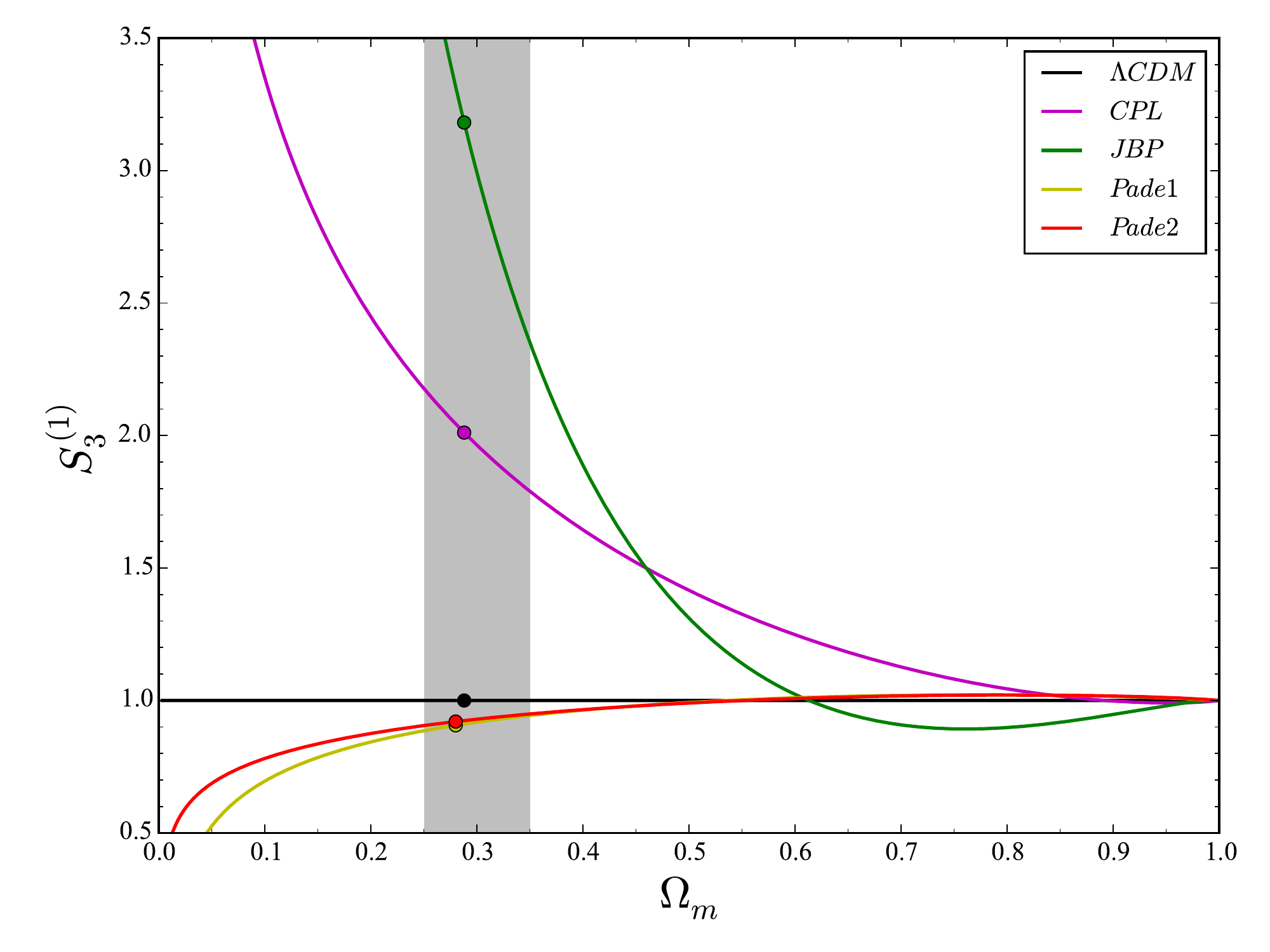}
\includegraphics[width=8cm,height=6cm]{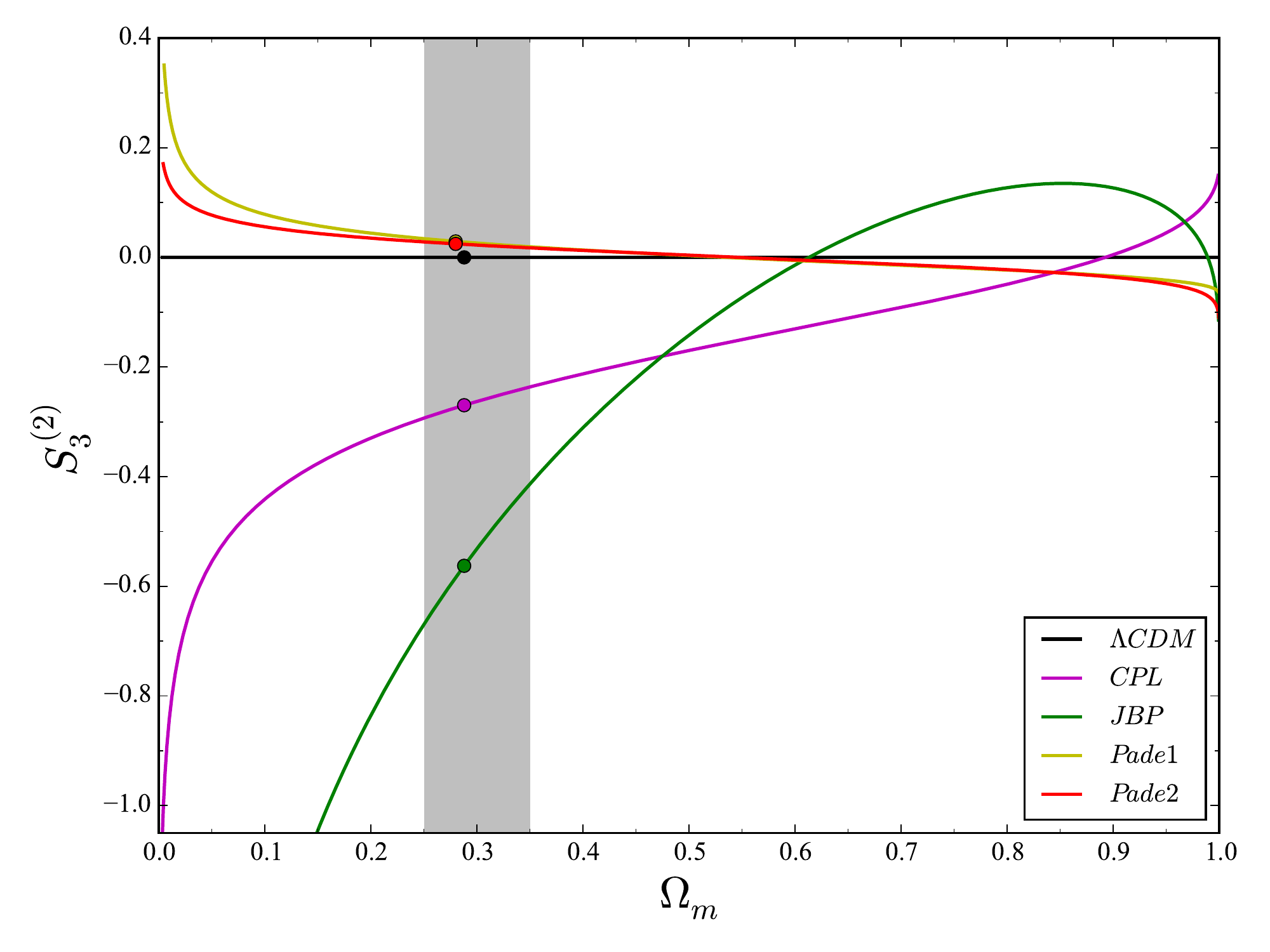}
\caption{The evolution trajectories of $S_3^{(1)}$ and $S_3^{(2)}$ with $\Omega_m$. Large values $\Omega_m\rightarrow 1$ represent the early universe ($z\gg 1$), while small values $\Omega_m \rightarrow 0$ correspond to the future universe ($z \rightarrow -1$). The vertical gray band centered at $\Omega_{m0}=0.3$ roughly stands for the present epoch. The points on line are the current values of corresponding model.}\label{F2}
\end{figure}

\begin{figure}
\centering
\includegraphics[width=8cm,height=6cm]{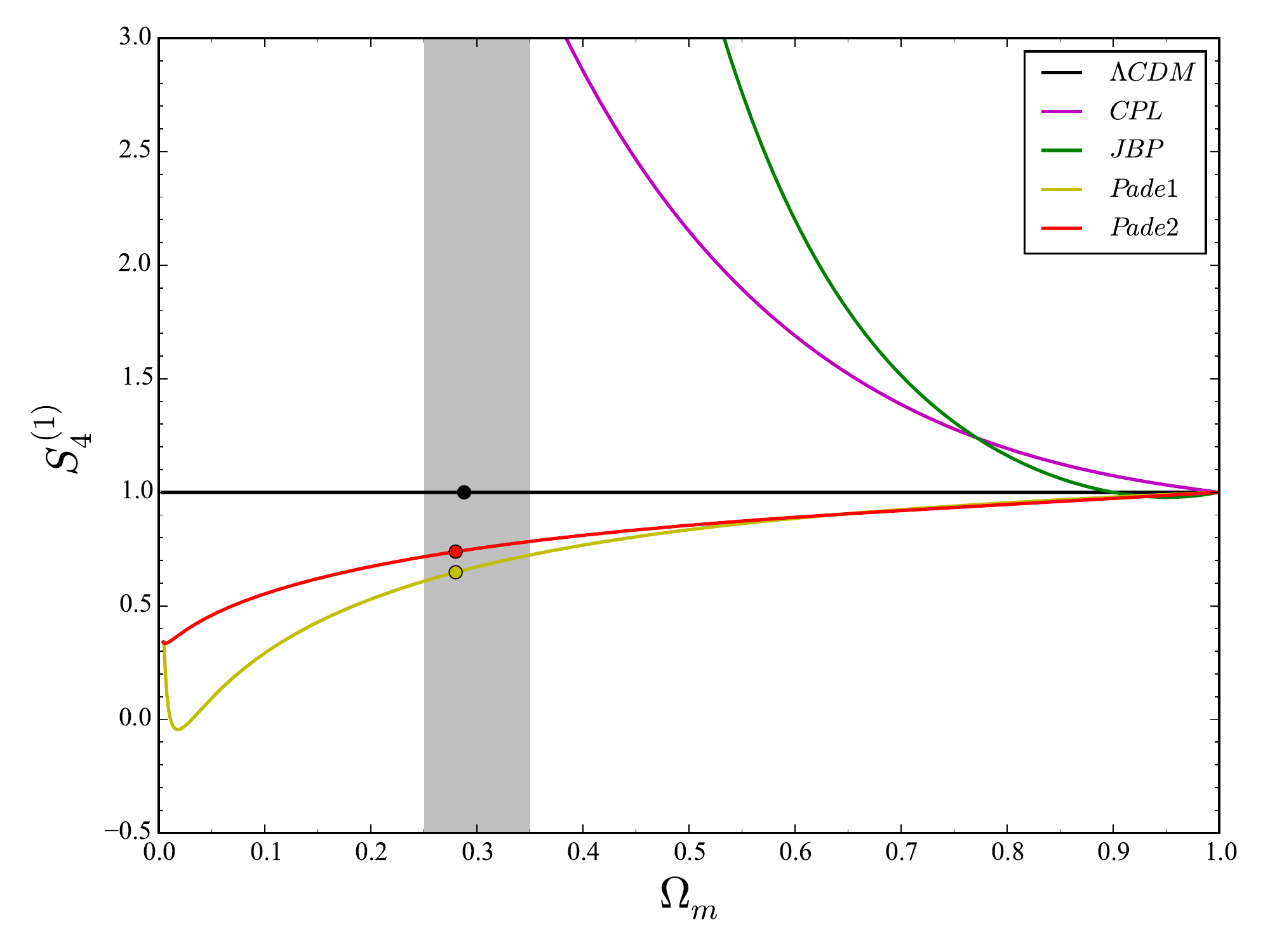}
\includegraphics[width=8cm,height=6cm]{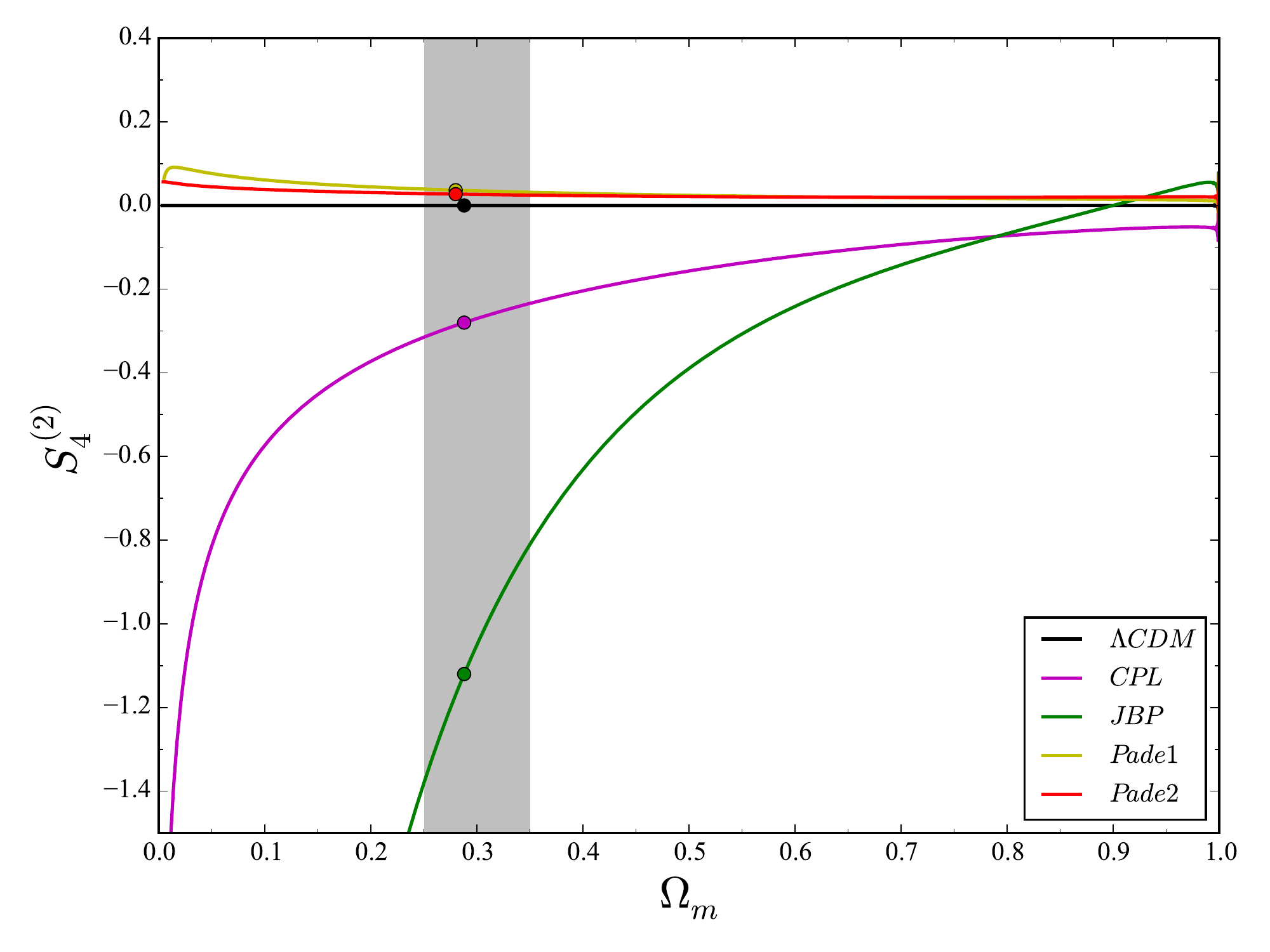}
\caption{The evolution trajectories of $S_4^{(1)}$ and $S_4^{(2)}$ with $\Omega_m$.}\label{FS4}
\end{figure}
The evolution trajectories of $S_3^{(1)}$ and $S_3^{(2)}$ with $\Omega_m$ are given in Fig. \ref{F2}. Large values $\Omega_m\rightarrow 1$ represent the early universe ($z\gg 1$), while small values $\Omega_m \rightarrow 0$ correspond to the future universe ($z \rightarrow -1$). It is obvious that the degeneracy among $\Lambda$CDM, CPL, JBP is perfectly broken in the present epoch. But the deviations between $\Lambda$CDM and Pad\'{e}(\uppercase\expandafter{\romannumeral1}), (\uppercase\expandafter{\romannumeral2}) are so small that they could not be distinguished. The discrimination of  $S_3^{(2)}$ is similar as that of $S_3^{(1)}$.

Due to $S_4$ as fourth derivative hierarchy is higher than $S_3$, it should be more powerful to distinguish models. As shown in Fig. \ref{FS4}, $\Omega_m-S_4^{(1)}$ can not only differentiate CPL, JBP, but also distinguish Pad\'{e}(\uppercase\expandafter{\romannumeral1}), (\uppercase\expandafter{\romannumeral2}) from $\Lambda$CDM. Although it could not differentiate Pad\'{e}(\uppercase\expandafter{\romannumeral1}) from (\uppercase\expandafter{\romannumeral2}) at present epoch, this can definitely be realized in the future. In perspective of transverse comparison, the discrimination of $S_4^{(1)}$ is better than $S_4^{(2)}$. Compared with $S_3^{(2)}$ in Fig. \ref{F2}, we find that the Statefinder $S_4^{(2)}$ does not bring obvious promotion of discrimination.
Naturally, $S_5$ and higher derivative hierarchy are expected to have more powerful discrimination. In the early universe, almost all these models are nearly degenerate into the $\Lambda$CDM model. In other words, the Statefinder hierarchy could not distinguish such models in high-redshift region. However, in the future, the evolutions of those models can be differentiated, especially in $S_4^{(1)}-\Omega_m$.

\begin{figure}
\centering
\includegraphics[width=8cm,height=6cm]{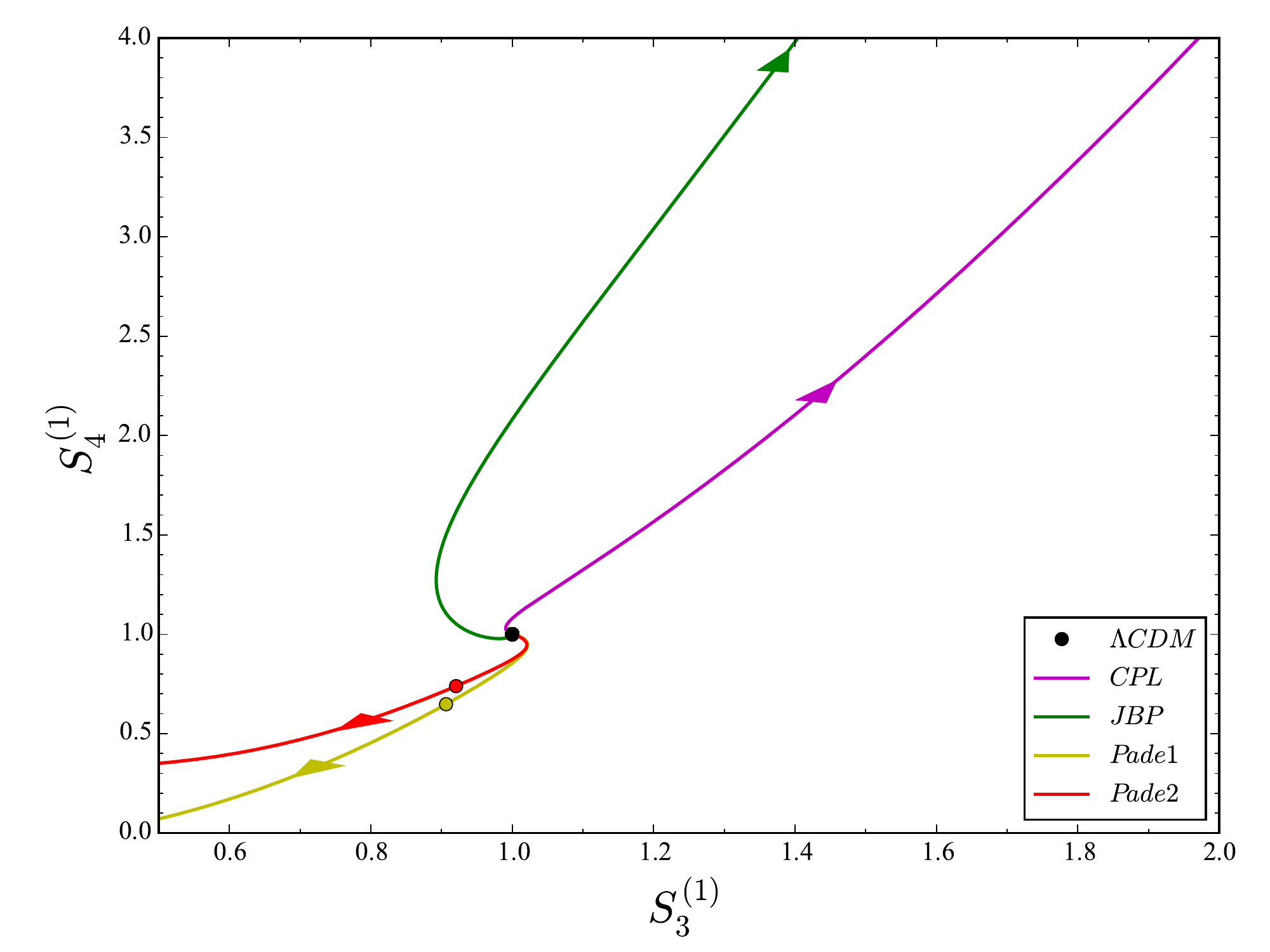}
\caption{The Statefinder $\{S_3^{(1)},S_4^{(1)}\}$ are shown for different parametrization dark energy models. The dots represent the current vaules and the arrows show time evolution. The current values of CPL and JBP are so large that they could not be shown in figure. }\label{F3}
\end{figure}

We plot the evolutionary trajectories of $\{S_3^{(1)},S_4^{(1)}\}$ to compare these models in Fig. \ref{F3}.
The result is similar to $S_4^{(1)}-\Omega_m$. The deviations of $\Lambda$CDM, CPL and JBP are quite large at present epoch, especially the current values of CPL and JBP are so large that they run over the figure, which indicates such models could be distinguished well. The Pad\'{e}(\uppercase\expandafter{\romannumeral1}) and Pad\'{e}(\uppercase\expandafter{\romannumeral2}) could be distinguished from $\Lambda$CDM at the present epoch but the discrimination between them will only appear in the future.

\begin{figure}
\centering
\includegraphics[width=8cm,height=6cm]{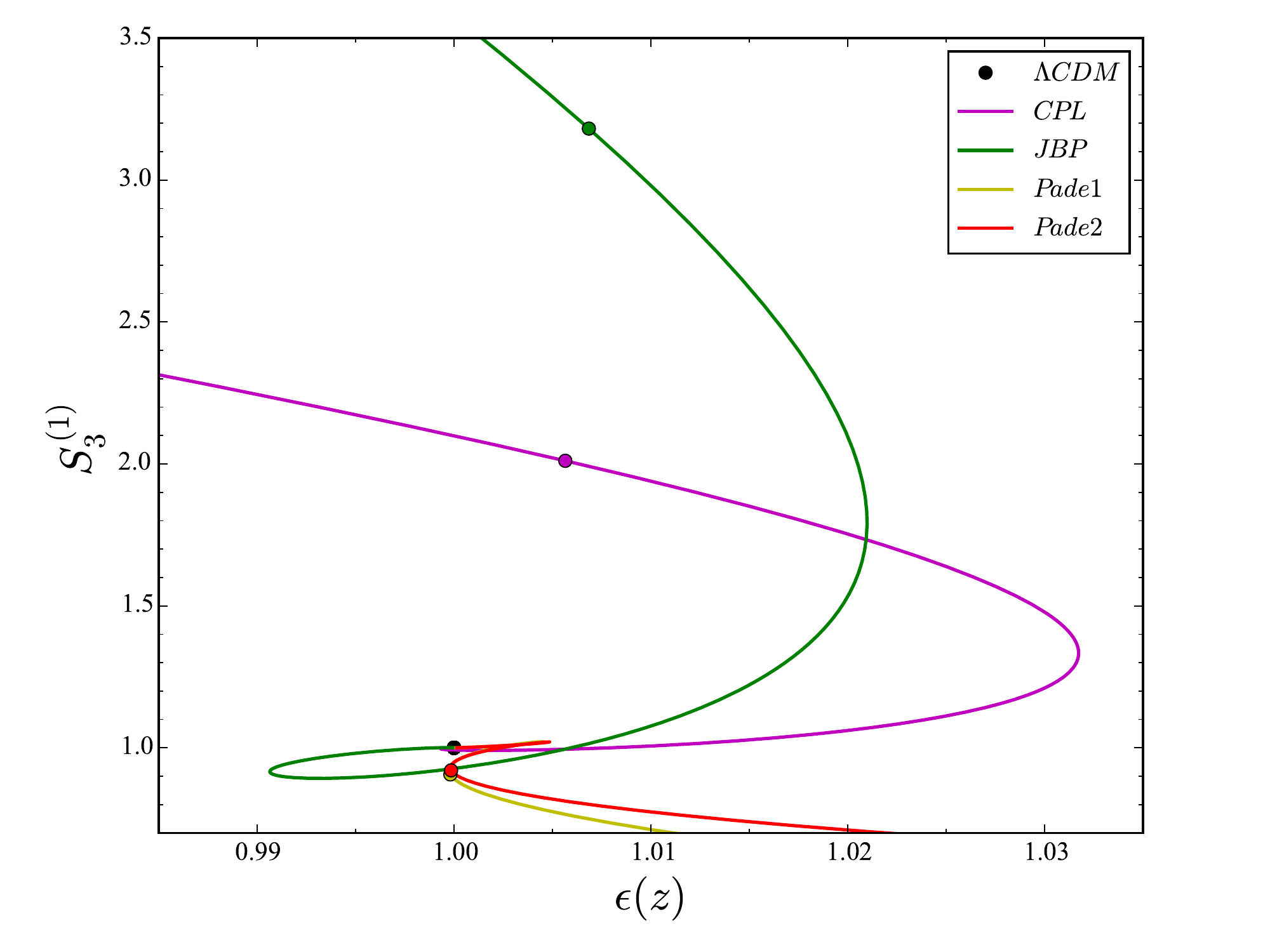}
\includegraphics[width=8cm,height=6cm]{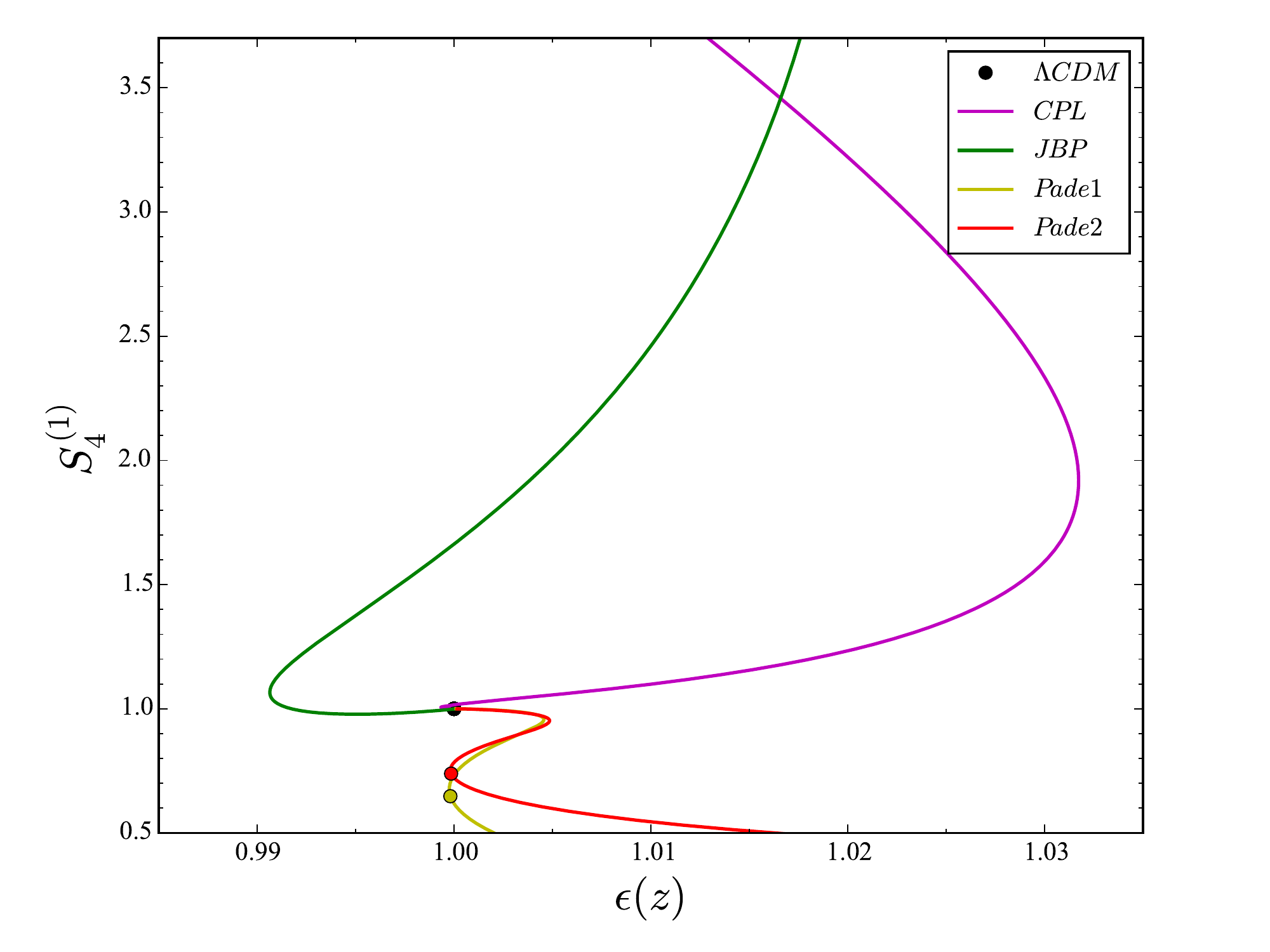}
\caption{The composite diagnostic  $\{\epsilon(z), S_3^{(1)}\}$ and $\{\epsilon(z), S_4^{(1)}\}$, respectively. The current values are marked by dots.}\label{F4}
\end{figure}

The composite null diagnostics $\{\epsilon(z), S_3^{(1)}\}$ and $\{\epsilon(z), S_4^{(1)}\}$ are also employed to distinguish these dark energy models, which is shown in Fig. \ref{F4}. In Refs. \citep{Arabsalmani:2011fz,Li:2014mua,zhang2014diagnosing,yu2015statefinder}, these diagnostics played a very important role in discriminating a number of dark energy models from $\Lambda$CDM.
In $\{\epsilon(z), S_3^{(1)}\}$ panel, CPL and JBP can separate from other models, while $\Lambda$CDM, Pad\'{e}(\uppercase\expandafter{\romannumeral1}) and Pad\'{e}(\uppercase\expandafter{\romannumeral2}) could not be discriminated at the present epoch. In $\{\epsilon(z), S_4^{(1)}\}$ panel, the result is better. The $\Lambda$CDM is differentiated from Pad\'{e} models, while Pad\'{e}(\uppercase\expandafter{\romannumeral1}) and (\uppercase\expandafter{\romannumeral2}) are still difficult to be distinguished between them at the present epoch. Thinking of Fig. \ref{F2}, the third and fourth order derivatives have given the similar effective results. This indicates that the supplement of Statefinders, the fractional growth parameter $\epsilon(z)$, does not play a significant role for discrimination of such parametrization dark energy models.

\section{Conclusions and discussions}\label{end}
In this paper, we explore the discrimination of some parametrization dark energy models including CPL, JBP, Pad\'{e}(\uppercase\expandafter{\romannumeral1}), (\uppercase\expandafter{\romannumeral2}) and $\Lambda$CDM. They are highly degenerate in $z-E(z)$ and $z-q(z)$, so they can not be distinguished in this scale. And then, we use Statefinder hierarchy and the growth rate of perturbations to discriminate such parametrization dark energy models. By using $S_3^{(m)}$, which contains third derivatives of $a(t)$, the CPL and JBP could be distinguished from $\Lambda$CDM and Pad\'{e}(\uppercase\expandafter{\romannumeral1}), (\uppercase\expandafter{\romannumeral2}), while the three models mentioned later can not be distinguished from each other. We also use the composite diagnostic $\{\epsilon(z), S_3^{(1)}\}$ to plot their trajectories. The result can not be improved obviously. Naturally, the higher order derivative $S_4^{(1)}$ has more powerful discrimination. It can not only differentiate CPL, JBP, but also distinguish Pad\'{e}(\uppercase\expandafter{\romannumeral1}), (\uppercase\expandafter{\romannumeral2}) from $\Lambda$CDM. Although it could not differentiate Pad\'{e}(\uppercase\expandafter{\romannumeral1}) from (\uppercase\expandafter{\romannumeral2}) at present epoch, this can definitely be realized in the future. Another Statefinder $S_4^{(2)}$ does bring degeneration instead of promotion of discrimination. The corresponding composite diagnostic $\{\epsilon(z), S_4^{(1)}\}$ has also been considered, while the results obtained are similar to $\{\Omega_m, S_4^{(1)}\}$. This indicates that the supplement of Statefinders, the fractional growth parameter $\epsilon(z)$, does not play a significant role for discrimination of such parametrization dark energy models. In conclusion, CPL, JBP, Pad\'{e} and $\Lambda$CDM could be discriminated at present epoch, and the discrimination of Pad\'{e}(\uppercase\expandafter{\romannumeral1}), (\uppercase\expandafter{\romannumeral2}) can only be realized in the future. Statefinder hierarchy is a pretty useful and effective method. Using this method, some parametrization dark energy models can be distinguished from $\Lambda$CDM or even from each other.

Now, let us review these parametric forms of EoS. As previously mentioned, the CPL can be regarded as the Taylor series expansion of EoS $w$ with respect to scale factor $a$ to first order. The second item of Eq. (\ref{cpl}) represents the deviate degree from $w_0$ ($\Lambda$CDM could be regarded as the special form of $w_0=-1$, so we consider the deviation of parametrization models from $\Lambda$CDM in the following). Similarly, the second item of Eq. (\ref{JBP}) can also be seen as a deviation from $\Lambda$CDM. Since the squared term in denominator, this deviation should be larger than the deviation between CPL and $\Lambda$CDM. These deviations could be expressed in evolution trajectories of Statefinder hierarchy, the deviation of JBP from $\Lambda$CDM is larger than CPL in any figure. In mathematics, a Pad\'{e} approximant is the best approximation of a function by a rational function of given order \citep{PP}. In addition, the Pad\'{e} approximant often gives better approximation of the function than truncating its Taylor series \citep{PP}. The Pad\'{e}(\uppercase\expandafter{\romannumeral1}), (\uppercase\expandafter{\romannumeral2}) are all based on the Pad\'{e} approximant. If we take $\Lambda$CDM as the standard, the superiority of Pad\'{e} parametrizations could be reflected in figure of Statefinder hierarchy because they are the most difficult to be distinguished from $\Lambda$CDM. However, on the other hand, the Pad\'{e} parametrizations with superiority have three free parameters which is one more than CPL and JBP.

\acknowledgments{
J.-Z. Qi would like to express his gratitude towards Prof. Rong-Jia Yang for his generous help. This work is supported by the National Natural Science Foundation of China (Grant Nos. 11235003, 11175019, and 11178007).

\bibliography{Notes}
\end{document}